# Amplification of nanosecond laser pulse chain via dynamic injection locking of laser diode

JUN HE[1,2,3], GANG JIN[1,2], BEI LIU[1,2], AND JUNMIN WANG[1,2,3,*]

[1] State Key Laboratory of Quantum Optics and Quantum Optics Devices, Shanxi University, Tai Yuan 030006, Shan Xi Province, China
[2] Institute of Opto-Electronics, Shanxi University, Tai Yuan 030006, Shan Xi Province, China
[3] Collaborative Innovation Center of Extreme Optics, Shanxi University, Tai Yuan 030006, Shan Xi Province, China
* Corresponding author: wwjjmm@sxu.edu.cn

**We report a novel optical pulse generation method for high-speed wavelength switching of amplified nanosecond (ns) laser pulses resonant to atomic transitions. Under free-running condition, a slave laser diode is blue-detuned with tens of GHz relative to the master laser. A nanosecond pulse chain generated by modulating the continuous-wave master laser with a fiber-pigtailed electro-optical intensity modulator is injected into the slave laser diode to fast switch the slave laser's wavelength back and forth. The output beam of slave laser is filtered by a temperature-controlled etalon to get the amplified pulse chain. Based on our dynamic injection locking scheme, we produce a ns-scale square pulse chain with an effective ON/OFF ratio ~$10^8$, considering at least the 60 dB scattering suppression by tuning light-atom interactions with far off-resonance detuning and 26.7 dB suppression ratio of the etalon. By studying the dynamic processes of injection locking, we determine the dependence of injection locking on both the injection power and the frequency detuning.**

Quantum computers and quantum information storage systems are built up from multiple microscopic physical systems. The set of criteria for quantum device operation requires control of light-matter interactions, including spatial matching or temporal matching, such as wavelength, single pulse energy and ON/OFF ratio [1,2]. In recent decades, single qubits have been used to perform fast, controlled excitation for single photon preparation [3,4]. Efficient coupling of a single pulse to a single atom is important to enable fast cycling [5,6,7]. A promising approach is free space coupling, where the coupling scheme avoids modification of the electromagnetic field density and offers greater design freedom [8,9]. Taking above consideration into account the pulse duration should be controlled not only satisfying the Fourier limitation but also satisfying the maximum Rabi oscillation frequency. For the manipulation of trapped atoms, we are interested in the situation where the interaction is active in ON state and perturbation is suppressed in OFF state.

An electro-optic intensity modulator (EOIM) is generally used to modify a continuous-wave laser beam to generate controlled ON/OFF pulses. However, while approaches based on EOIMs are promising for ns or ps pulse production, it is difficult to achieve higher ON/OFF ratios. Due to technique limitations, the ON/OFF ratio is approximately 20 dB. Some approaches add EOIMs in series to increase the ON/OFF ratio. Kosen [10] and Leong *et al.* [11] achieved a ratio of ~40 dB for amplitude modulation using two EOIMs in series. Conventional approaches involve amplifying the output power using a tapered amplifier [12]. Because the tapered amplifier works in continuous mode, it is sensitive to the pulse modes. It therefore appears difficult to achieve a low repetition rate. Another more serious problem is that the amplified spontaneous emission of the tapered amplifier is always present, which increases multi-photon excitation in the OFF state [13].

Another alternative approach is injection locking technology [14]. Using a distributed feedback (DFB) diode laser as a slave laser, the microcavity can respond rapidly to ns or ps pulses. Due to the extreme frequency sensitivities of interaction between atom and light, it is possible to reduce the resonance interaction by switching the wavelength between the ON and OFF states. In this paper, we expand the power-threshold-dependent injection locking, presenting a wavelength switching between the ON and OFF states, and the power amplifier on the ON state and power suppression on OFF state. Under the conditions of far off-resonance and filter, the scattering rate shows a difference of approximately 8 orders of magnitude for the ON and OFF states (effective ON/OFF ratio).

Gain competition often inhibits the simultaneous lasing of multiple longitudinal modes. Consider a mode of a slave laser that is matched to a mode of the master laser. For successful operation of the slave laser, it is essential to increase the carrier consumption that arises from the master laser mode over a short period [15]. To achieve rapid injection locking, the cavity of slave laser should be short enough to ensure steady state performance in the presence of the master laser pulses. The electric field of the slave laser ( $E_S = \sqrt{P_S(\mathrm{t})}\exp[\mathrm{i}\omega_s \mathrm{t} + \phi_0(\mathrm{t})]$ ) and the master laser ( $E_M = \sqrt{P_M(\mathrm{t})}\exp[\mathrm{i}\omega_M \mathrm{t} + \phi(\mathrm{t})]$ ) can be described as classic electromagnetic field, where $\omega$ is the frequency, $\varphi$ is the phase, and $P$ is the amplitude of the electromagnetic field. In the steady state approximation, the properties of the master laser mode should match those of the corresponding slave laser mode, which can be described by $E_i = \sqrt{P_i(\mathrm{t})}\exp[\mathrm{i}(\omega_i + \phi_i(\mathrm{t}))]$.

When the slave laser runs freely, a competing architecture then exists for production of a single-frequency laser. Here we consider the freely running mode and the injection mode. The gain coefficients and loss coefficients are $G_i$ and $\delta_i$, or $G_S$ and $\delta_S$, corresponding to the injection mode and the free running mode of the slave laser, respectively. α is the linewidth gain coefficient. The time-dependent dynamic equations for the electric fields are then

$$\frac{dE_S(t)}{dt} = \frac{1}{2}(1-i\alpha)(G_S - \delta_S)E_S(t) \tag{1}$$

$$\frac{dE_i(t)}{dt} = \frac{1}{2}(1-i\alpha)(G_i - \delta_i)E_i(t) \tag{2}$$

$$\frac{dN(t)}{dt} = J - \frac{N(t)}{\tau_s} - G_S E_S(t) - G_{SS}E_{SS}(t) \tag{3}$$

Based on the rate equation, the dynamic response of the injection mode is described using

$$\frac{dP_S}{dt} = (G_S - \delta_S)P_S(t) \tag{4}$$

$$\frac{d\phi_S(t)}{dt} = \frac{\alpha}{2}(G_S - \delta_S) \tag{5}$$

$$\frac{dP_{SS}(t)}{dt} = (G_{SS} - \delta_{SS})P_S(t) + 2\gamma_C\sqrt{P_i(t)P_S(t)}\cos\phi_{SS} \tag{6}$$

$$\frac{d\phi_i(t)}{dt} = \Delta\omega + \frac{\alpha}{2}(G_i - \delta_i) - \gamma_C\sqrt{\frac{P_M(t)}{P_S(t)}}\sin\phi_i \tag{7}$$

$$\frac{dN(t)}{dt} = J - \frac{N(t)}{\tau_S} - G_S E_S(t) - G_i E_i(t) \tag{8}$$

In the steady state, $P_S(t) = |E_S(t)|^2 = E_S^*(t)E_S(t)$, and thus

$$\frac{dP_{SS}(t)}{dt} = 0 \tag{9}$$

$$\frac{d\phi_{SS}(t)}{dt} = 0 \tag{10}$$

Combining (9) and (10) yields the injection frequency range, as follows:

$$\Delta\omega = \gamma_C\sqrt{\frac{P_i}{P_S}}(\alpha\cos\phi_{ss} + \sin\phi_{ss}) = \gamma_C\sqrt{\frac{P_i}{P_S}}\sqrt{1+\alpha^2}\sin(\phi_{ss} + arctg\alpha) \tag{11}$$

To build the oscillation above the injection locking threshold under far off-resonance conditions relative to the injection laser frequency, the phase condition in Equation (11) is $-\pi/2 \le \phi_{SS} \le \pi/2$, which corresponds to $(-\pi/2 + arctg\alpha) \le (\phi_{SS} + arctg\alpha) \le (\pi/2 + arctg\alpha)$. Therefore, the boundary conditions from Equation (11) are

$$-\gamma_C\sqrt{\frac{P_i}{P_S}} \le \Delta\omega \le \gamma_C\sqrt{\frac{P_i}{P_S}}\sqrt{1+\alpha^2} \tag{12}$$

where the boundary conditions determine the linewidth gain coefficient α and the optical coupling coefficient $\gamma_C$. The mode oscillations depend on the injected power, which has an increasing function with time. The injection locking frequency range depends on both the α and $\gamma_C$.

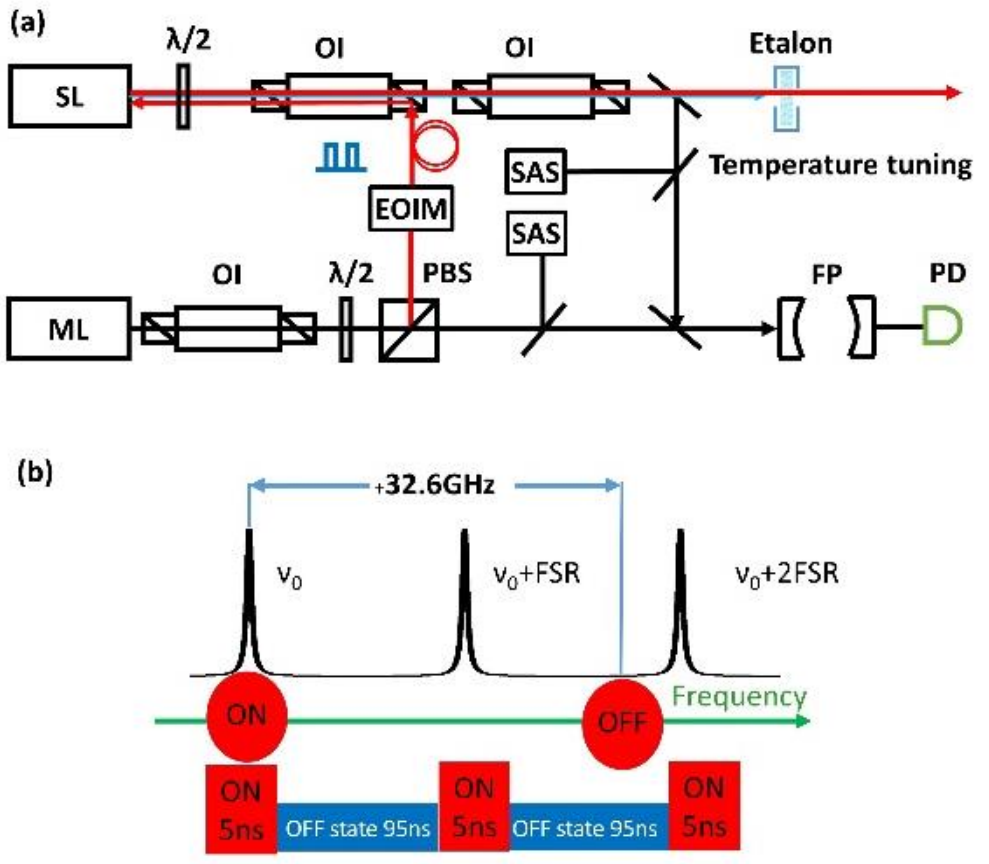

**Fig. 1.** (Color online) (a) Schematic of experimental setup, where ML and SL are the master laser and the slave laser, respectively. IO: optical isolator. EIOM: electro-optical intensity modulator; SAS: saturation absorption spectroscope; FP: Fabry-Pérot cavity; PD: photodiode; λ/2: a wave plate; PBS: a polarizing beam splitter. (b) Dynamic injection locking of laser diode using detuning of +32.6 GHz. A temperature-controlled etalon filters the output beam of slave laser.

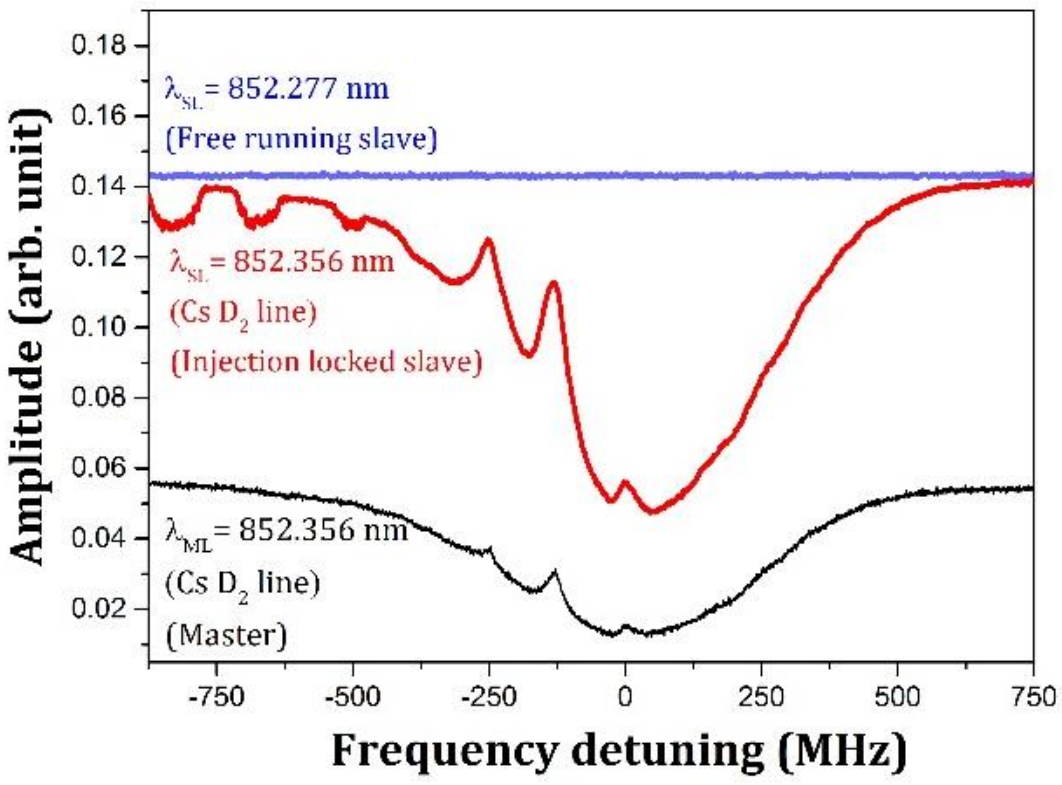

**Fig. 2.** (Color online) Implementation of fast injection locking on ns time scale, with the SAS operating under freely running and injection locked conditions. The spectrum of atoms demonstrates the frequency locking between master laser and slave laser. When the master laser frequency is continuously changing, the spectrum of the slave laser shows stable injection locking without mode hopping, this corresponds to the locked range being more than 1 GHz,

Instead of direct modulation approach, we inject the EOIM output into the slave laser. In our experiment, the master laser beam is output from a diode laser (DL100, Toptica). As shown in Fig.1 (a), the output beam passes through an optical isolator (OI) and is then split into two parts. One of the beams passes through the cesium cell (not shown), which is used to monitor the frequency of the master laser. The other beam passes through the EOIM to generate the ns pulses. The EOIM is driven by a waveform generator (AVM-1-C-P-PN-AK1, AVTECH). The EOIM output pulses are coupled in the slave laser using an optimal OI. A temperature-controlled etalon is used to filter the output beam of slave laser, as shown in Fig. 1 (b). A 149 mm confocal Fabry-Pérot (FP) cavity is used to monitor two laser modes. In the experiments, the master laser is locked to the $6S_{1/2}$ ($F_g = 4$) - $6P_{3/2}$ ($F_e$ = 5) transitions by the polarization spectrum. When the injected condition is satisfied, the spectrum of the slave laser is the same as that of the master laser, and thus corresponds to the same laser mode Fig. 2. When the master laser frequency is continuously changing, the spectrum of the slave laser shows stable injection locking without mode hopping, this corresponds to the locked range being more than 1 GHz.

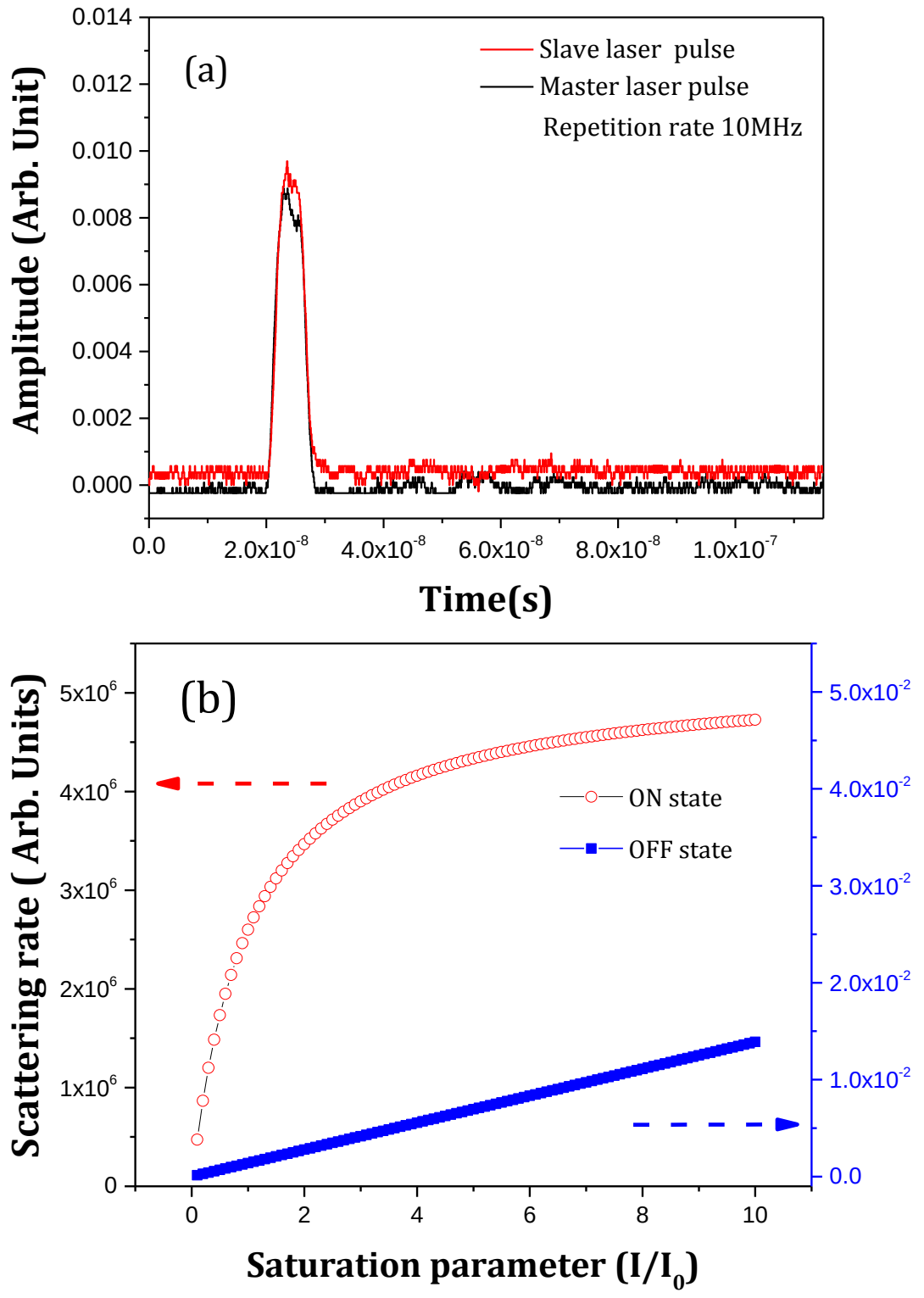

**Fig. 3.** (Color online) (a) Temporal pulses for the master laser and the slave laser in detail. The slave laser is running freely with detuning of +32 GHz. (b) Effective scattering rate versus saturation parameter, considering 26.7 dB suppression ratio of the etalon. The red circles and blue squares denote the simulation of scattering rates under on-resonance conditions and under far off-resonance conditions with detuning of +32 GHz, respectively. The scattering rate depends on the saturation parameter.

We use the etalon to filter the output beam. The transmission is varied periodically with the laser frequency by controlling the operating temperature. The free spectral range is 18.4 GHz, and bandwidth is 0.8 GHz. The temperature matching range is approximately 6.61°C and tuning rate 2.772GHz/℃. The peak transmission is ~91.7% and the suppression ratio is 26.7 dB. The power from the slave output after the etalon is ~50 mW. To achieve the injection locking, the frequency of the slave laser is typically maintained at near-resonance relative to the master laser frequency [16]. However, the resonance condition implies that the freely running frequency is the nearly same when compared with the locked phase, which results in difficulty for filtering the laser beam in the OFF state. The frequency condition is expected to be far off-resonance relative to the master laser or the injection locking state (ON state). This is usually done by increasing the laser current such that the slave laser will tune monotonically with a characteristic slope of approximately 0.002 nm/mA. Mode hops typically occur with a free spectral bandwidth of up to 0.15 nm. Injection locking is monitored using an FP cavity under various operating currents. By careful choice of both the current and temperature parameters, the slave laser is run under condition of detuning of +32.6 GHz. Under ns injected pulse conditions, the slave laser shows a similar pulse mode that follows that of the master laser, as shown in Fig. 3(a). When compared with the limited ON/OFF ratio of near-resonance injection locking, injection locking with far detuning shows a better pulse profile. We are more interested in the scattering rate of the ON/OFF state. As shown in Fig. 3(b), under the conditions of the +32GHz far off-resonance the scattering rate shows a difference of ~8 orders of magnitude in ON/OFF state, considering at least the 60 dB scattering suppression and 26.7 dB suppression ratio of the etalon.

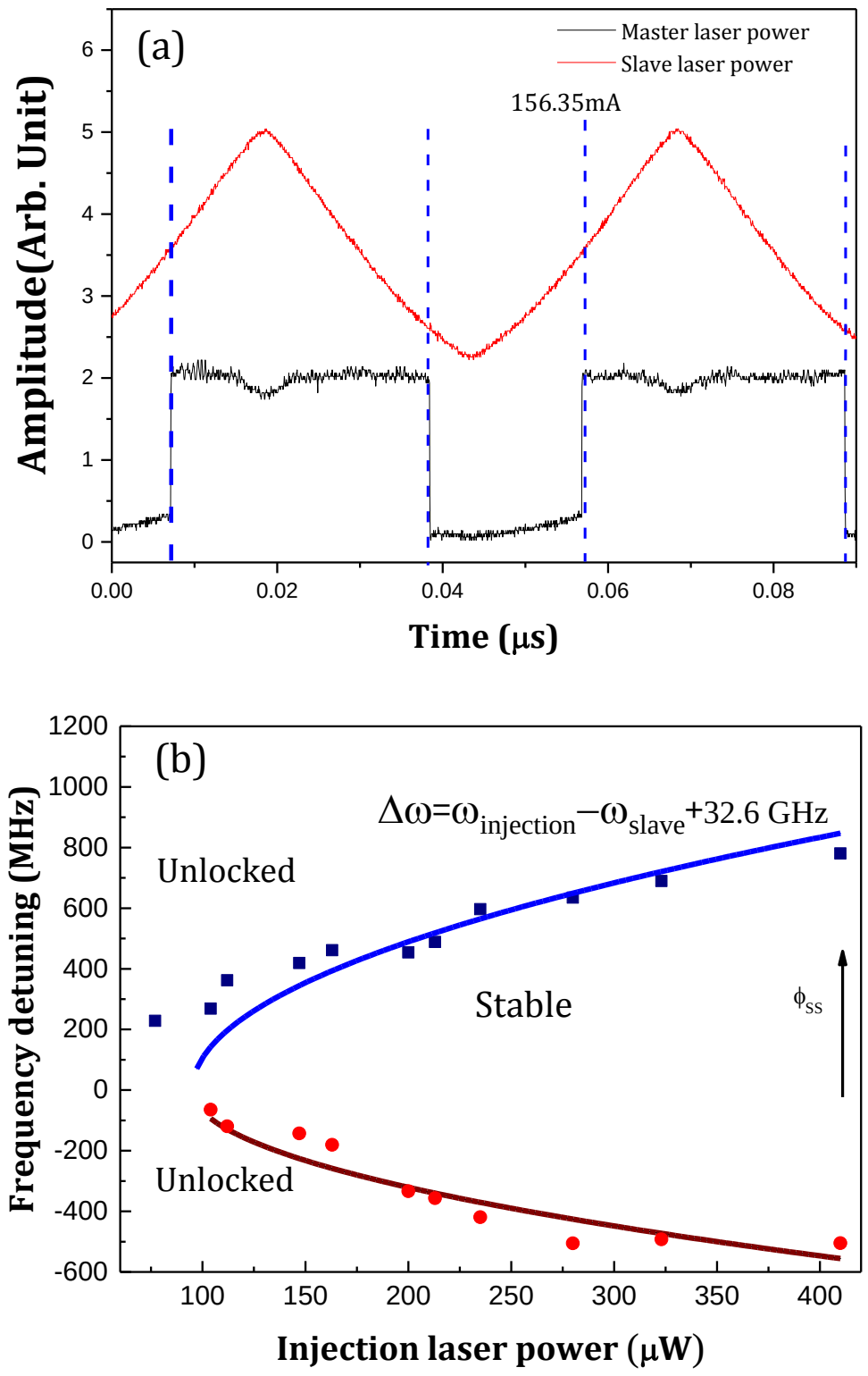

**Fig. 4.** (Color online) Dependence of injection frequency range on injected light power. (a) With linearly increasing master laser power, the output power of the slave laser changes with a rectangular profile, which corresponds to a typical threshold. (b) Red dots and purple squares represent the different detunings of the frequency. The data are fitted using equations (6, 7, 8), as indicated by the red and blue lines.

The injection locking depends on the input laser power threshold. Using the arbitrary waveform generator with optimized programmable software, we drive the EOIM to generate a triangle wave for linear scaling of the output power. By scaling the injected power, we observe the threshold dependence of injection locking of the slave laser. The higher power output represents the injection phase, and the lower power output represents the freely running phase, as shown in Fig. 4(a). The multimode oscillation is an increasing function, where $P_{ss}$ increases with increasing injection time. The injection range frequency boundary is dependent on the α and $\gamma_c$ [17].We measure the frequency dependence using the FP cavity, and also measure the dependence of the frequency on the injected power, as shown in Fig. 4(b). Fitting of Equation (11) to the data yields the $\gamma_c = 2\pi \times 8370$ MHz and $\alpha = 1.15$. The observed injection power threshold is approximately 95 μW. The dominant errors come from the imperfections in the injection power measurement process. Experimental data are combined with a simulation based on Equations (6), (7) and (8), while taking the optical coupling coefficient into account, and give a minimum injection time of 19 ps. In our experiments, we achieved ns-scale injection locking that is limited by the waveform generation process. We also observe optical bistability. When the injection power is linearly increased or reduced, the threshold varies, and it is dependent on the optical injection frequency. A similar phenomenon can also be observed in Fabry-Pérot laser diodes [18].

In conclusion, we present a robust but simple scheme to generate ns-scale pulse chain with an effective ON/OFF ratio $\sim 10^8$, considering at least the 60 dB scattering suppression by tuning interacting of light with atoms in far off-resonance detuning and 26.7 dB suppression ratio of the etalon. The single-pulse energy of the nanosecond laser pulse is amplified via dynamic injection locking process. By studying the dynamic processes of injection locking, we determine the dependence of injection locking on both the injection power and the frequency detuning. The proposed injection locking scheme is not insensitive to the ON/OFF ratio of the injected light. For the practical output from EOIM of master laser, the maximum peak pulse power of is 1.25 mW, the average pulse power is 62.5 μW, and the single-pulse energy is 6.25 pJ. For the slave output laser beam after the etalon, the peak pulse power is ~50 mW, the average pulse power is 0.25 mW, and the single-pulse energy is 0.25 nJ, which is sufficient to achieve several π-pulses for Rabi oscillation of neutral atom. The gain coefficient of the single-pulse energy is approximately 526. There is also the possibility of using higher power slave lasers to achieve higher peak powers.

The most important result is that under condition of injection locking with a wavelength switching between the ON and OFF states, the scattering rate in OFF state can be substantially reduced. The combination of a short pulse with high power and a high ON/OFF ratio can provide powerful temporal matching for light-matter interaction on the single particle level. These findings will also be beneficial for other applications, including random number generation [14] and quantum key distribution [19], and also may be applied in laser pulse repeaters that are used at regular intervals to regenerate and reshape optical pulses.

**Funding.** The National Natural Science Foundation of China (Grant Nos. 11274213, 61475091, 61205215, and 61227902), the National Major Scientific Research Program (973 Project) of China (Grant No. 2012CB921601).